\documentclass[pre,twocolumn,groupedaddress,showpacs,showkeys,amsmath]{revtex4}

\usepackage{graphicx}
\usepackage{dcolumn}
\usepackage{bm}

\begin{document}


\title{Analysis of the loop length distribution for the 
negative weight percolation problem in dimensions $d=2$ through $6$}

\author{G. Claussen}
\email{gunnar.claussen@uni-oldenburg.de}
\author{L. Apolo$^{1,2}$}
\email{lapolo00@ccny.cuny.edu}
\author{O. Melchert$^1$}
\email{oliver.melchert@uni-oldenburg.de}
\author{A. K. Hartmann$^1$}
\email{alexander.hartmann@uni-oldenburg.de}
\affiliation{
$^1$ Institut f\"ur Physik, Universit\"at Oldenburg, Carl-von-Ossietzky Strasse, 26111 Oldenburg, Germany\\
$^2$ City College of the City University of New York, New York, New York 10031, USA
}

\date{\today}


\begin{abstract}
We consider the negative weight percolation (NWP) problem 
on hypercubic lattice graphs with fully
periodic boundary conditions in all relevant dimensions 
from $d=2$ to the upper critical dimension $d=6$. The problem
exhibits edge weights drawn from disorder distributions that allow for weights
of either sign. We are interested in in the full ensemble of loops 
with negative weight, i.e.\ non-trivial (system spanning) loops
as well as  topologically trivial (``small'') loops. 
The NWP phenomenon refers to the disorder driven proliferation of system spanning loops of total negative 
weight. While previous studies where focused on the latter loops, we
here put under scrutiny the ensemble of small loops. Our aim is to 
characterize -using this extensive and exhaustive numerical study- the 
loop length distribution of the small loops right at and below the critical 
point of the hypercubic setups by means of two independent critical 
exponents. These can further be related to the results of previous 
finite-size scaling analyses carried out for the system spanning loops.
For the numerical simulations we employed
 a mapping of the NWP model to a 
combinatorial optimization problem that can be solved exactly by using
sophisticated matching algorithms. This allowed us to study
here numerically exact very large systems with high statistics.
\end{abstract} 

\pacs{}
\maketitle

\section{Introduction \label{sect:introduction}}

The statistical properties of lattice-path models on graphs,
equipped with quenched disorder, have experienced much attention during the 
last decades.
They have proven to be useful in order to characterize, e.g.,
linear polymers in disordered/random media \cite{kremer1981,kardar1987,derrida1990,grassberger1993,parshani2009}, 
vortex loops in high-$T_c$ superconductivity at zero field \cite{nguyen1998,nguyen1999,pfeiffer2002,pfeiffer2003}
and the $d=3$ XY model \cite{kajantie2000,camarda2006}, networks of vortex strings
found after a symmetry-breaking phase transition in field theories \cite{antunes1998,hindmarsch1995,strobl1997},
as well as
domain wall excitations in disordered media such as spin glasses \cite{cieplak1994,melchert2007} and 
the solid-on-solid model \cite{schwarz2009}. 
The precise computation of these paths can often be formulated in terms
of a combinatorial optimization problem and hence might allow for the
application of exact optimization algorithms 
\cite{papadimitriou1998} developed in computer science.

For an analysis of the statistical properties of these lattice path models, 
geometric observables and scaling concepts similar to those developed in 
percolation theory \cite{stauffer1979,stauffer1994,schakel2001} have been used conveniently. 
In the past decades, a large number of percolation problems in various
contexts have been investigated through numerical simulations.
Among those are problems, where the fundamental entities are string-like, 
similar to the lattice-path models mentioned in the beginning, 
rather than clusters consisting of occupied nearest neighbor sites 
as in the case of usual random bond percolation.

In a sequence of recent articles we have introduced 
\cite{melchert2008} and
investigated (see below) \emph{ the negative-weight percolation} (NWP),
a problem with subtle differences as compared to other string-like percolation
problems.
In the most basic NWP setup, one considers a regular lattice graph with periodic boundary conditions
(BCs), where adjacent sites are joined by undirected edges. Weights are
assigned to the edges, representing quenched random variables drawn from 
a distribution that allows for edge weights of either sign. The properties
of the weight distribution are further controlled by a tunable disorder 
parameter, signified $\rho$. For a given realization of the disorder, one then computes a 
configuration of loops, i.e.\ closed paths on the lattice graph, 
such that the sum of the edge weights that build up the loops is minimal 
and negative.
As an additional optimization constraint we impose the condition that 
the loops are not allowed to intersect; consequently there is no definition
of clusters in the NWP model. 
Regardless of the spacial dimension of the underlying (hypercubic) lattice
graph, the observables are always line-like, i.e.\ have an intrinsic dimension 
of $d=1$. Nevertheless, the loops may be fractal with
fractal dimensions $d_f>1$, see Ref.\ \cite{melchert2010a}.

\begin{figure}[t!]
\centerline{
\includegraphics[width=1.0\linewidth]{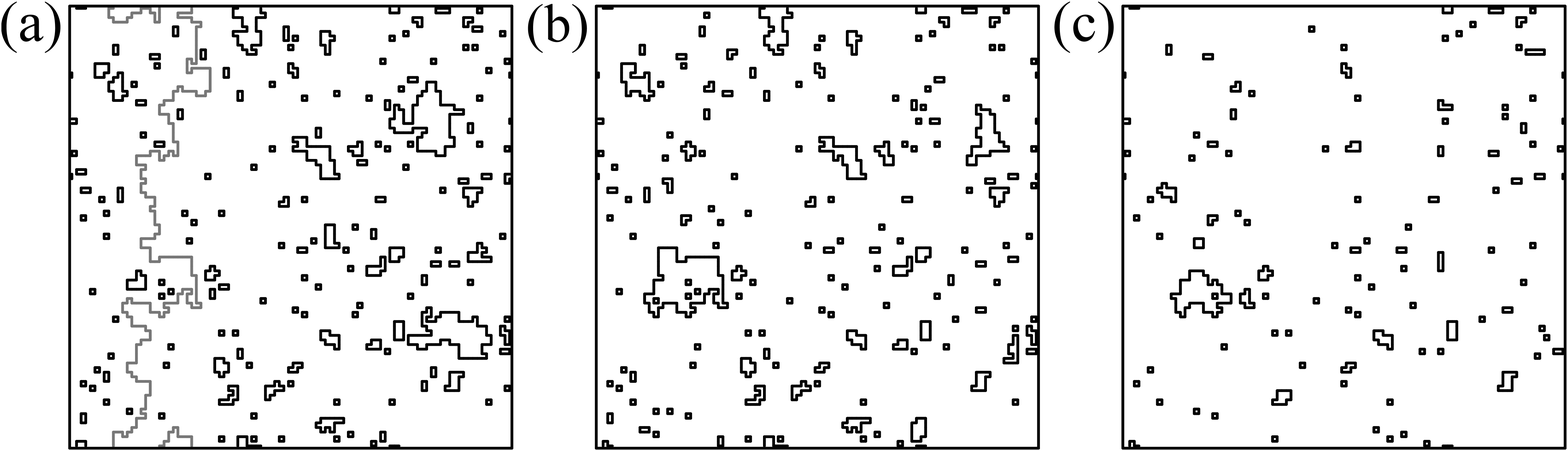}}
\caption{
Samples of minimum weight configurations of
loops for a $2D$ square lattice 
with side length $L\!=\!64$ and fully periodic boundary conditions.
The snapshots relate to different values of the
disorder parameter $\rho$, where (a) $\rho\approx\rho_c$, 
(b) $\rho^\prime < \rho_c$, and, 
(c) $\rho^{\prime \prime}<\rho^{\prime}$.
In the limit of large system sizes and above the critical point 
$\rho_c$, loops might span the lattice along at least one direction as,
 e.g., the gray
loop in (a).
For small values of $\rho$, loops with a comparatively large length appear to be suppressed exponentially.
\label{fig:samples2D}}
\end{figure}

The problem of finding these loops numerically
can be cast into a minimum-weight
path (MWP) problem, outlined in sect.\ \ref{sect:model} in more detail.
A pivotal observation is that, as a function of the disorder parameter 
$\rho$,  
the NWP model features a disorder driven, geometric phase transition,
\cite{melchert2008,apolo2009,melchert2010a} 
triggered by a vital change of the typical loop size (as discussed below in more detail).
In this regard, depending on the precise lattice setup and on the value of $\rho$, 
one can identify two different phases: 
(i)  a phase where the loops are ``small'', meaning that the linear extensions of
     the loops are small in comparison to the system size, see Figs.\ \ref{fig:samples2D}(b-c)
     (therein, the linear extension of a loop refers to its projection onto the independent 
     lattice axes), and,
(ii) a phase where ``large'' loops exist that span the entire lattice, see Fig.\ \ref{fig:samples2D}(a).
Regarding these two phases and in the limit of large system sizes, there is a particular 
value of the disorder parameter, signified as $\rho_c$, at which system spanning (or ``percolating'') loops appear 
for the first time.

Previously, we have investigated the NWP phenomenon for $2D$ lattice graphs \cite{melchert2008}
using finite-size scaling (FSS) analyses, where we characterized the underlying transition by 
means of a set of critical exponents. 
Considering different disorder distributions and lattice geometries, the exponents where found 
to be universal in $2D$ and clearly distinct from those describing other percolation phenomena.
In a subsequent study we investigated the effect of dilution on the critical 
properties of the $2D$ NWP phenomenon \cite{apolo2009}. Therefore we performed
FSS analyses to probe critical points along the critical line in the disorder-dilution plane 
that separates domains that exhibit or do not exhibit
 system spanning loops. 
One conclusion of that study was that bond dilution changes the universality class 
of the NWP problem. Further we found that, for bond-diluted lattices prepared
at the percolation threshold of $2D$ random percolation and at full disorder, the 
geometric properties of the system spanning loops compare well to those of ordinary 
self-avoiding walks.
We performed further simulations for the NWP model on hypercubic lattice graphs in dimensions
$d\!=\!2$ through $7$ \cite{melchert2010a}, where we found evidence for an upper critical 
dimension $d_u\!=\!6$ of the NWP phenomenon. This result was based on monitoring the critical 
exponents related to the NWP transition (one expects them to stay fixed
for $d\geq d_u$). 
We also studied numerically as well as analytically 
a variant of the NWP transition on 
$3$-regular random graphs (RRGs), i.e.\ graphs 
where each node has exactly $3$ neighbors and where there is no regular 
lattice structure. Hence, we obtained
direct access to the mean-field exponents that govern the model 
for $d\geq d_u$.  We obtained excellent agreement between
numerical and analytic results and
 could provide further support for the claim $d_u=6$.

All of the studies mentioned above where focused on the statistical properties of the 
largest loop (or more precisely, the \emph{longest} loop) for a given realization of the disorder
and the critical properties of the NWP model that derive from an analysis of 
these loops in the vicinity of the critical point $\rho_c$. 
Up to now, limited attention was payed to the ensemble of ``small'' loops that actually 
comprise the major part of loop segments in the vicinity of $\rho_c$.
As we found earlier, at this critical point the loops are rather isolated and well separated 
from each other, resembling a dilute gas of loops (cf.\ Fig.\ \ref{fig:samples2D}).
Further, the normalized and 
ensemble-averaged probability mass function (pmf) $n_\ell$ of 
loops having length $\ell$ right at $\rho_c$ was studied in 
Refs.\ \cite{melchert2008,melchert2010a}. It exhibits an algebraic 
decay similar to the distribution of cluster sizes at the 
critical point in ordinary random percolation \cite{stauffer1979,stauffer1994}, i.e.\ 
\begin{equation} 
n_\ell(\rho_c) \propto \ell^{-\tau} \quad \text{at} \quad \rho=\rho_c.\label{eq:loopDistribCrit}
\end{equation}
The numerical values of the decay exponent $\tau$ (also termed ``Fisher exponent'') found for 
the NWP model in $d=2$ through $7$ are listed in Tab.\ \ref{tab:tab1}.
Note that the Fisher exponent is only one out of two 
independent exponents that characterize the whole ensemble of loops.

%
\begin{table}[b!]
\caption{\label{tab:tab1} Critical properties that characterize the
NWP phenomenon  in $d\!=\!2\ldots6$.  From left to right:  Lattice
dimension $d$, critical point $\rho_c$, product $\nu_{\rm p} \cdot
d_{f,{\rm p}}$ of the critical exponents $\nu_{\rm p}$ and $d_{f,{\rm
p}}$  that describe the divergence of a typical length scale and the
fractal dimension $d_{f,{\rm p}}$ (measured at $\rho_c$),
respectively, as well as the length fluctuation exponent $\gamma_{\rm
p}$. Further, the table lists the Fisher exponent $\tau$ and the
loop-length cut-off exponent $\sigma$.  Note
that the figures in all but the  two right columns
 are taken from Ref.\
\cite{melchert2010a}.  The last column is meant to
check the scaling relation $\gamma_{\rm p}=(3-\tau)/\sigma$.  
The additional subscript {\rm p} indicates that
these exponents result from an analysis of the percolating loops.  }
\begin{ruledtabular}
\begin{tabular}[c]{llllllll}
$d$  & $L$ & $\rho_c$ & $\nu_{\rm p}d_{f,{\rm p}}$& $\gamma_{\rm p}$ & $\tau$ & $\sigma$ & $(3-\tau)/\sigma$ \\
\hline
2 & 512   & 0.340(1)  & 0.53(3) & 0.77(7)  & 2.59(3) & 0.53(3)& 0.77(10) \\
3 & 64    & 0.1273(3) & 0.69(2) & -0.09(3) & 3.07(1) & 0.71(1)& -0.10(1)\\
4 & 21    & 0.0640(2) & 0.78(3) & -0.66(5) & 3.55(2) & 0.78(2)& -0.71(1)\\
5 & 12    & 0.0385(2) & 0.86(4) & -1.06(7) & 3.86(3) & 0.88(2)& -0.98(1)\\
6 & 6     & 0.0265(2) & 1.00(3) & -0.99(3) & 4.00(2) & 0.97(4)& -1.03(2)\\ 
\end{tabular}
\end{ruledtabular}
\end{table}

In the present 
article, the second critical exponent that characterizes
the ensemble of small loops is addressed.
In this regard, the present article discusses the pmf 
$n_\ell(\rho)$ as
function of the disorder parameter $\rho$.
Consequently, the numerical effort to obtain these distributions,
in several dimensions $d=2,\ldots,6$ was much larger, compared
to the previous studies where the distribution was obtained just 
for $\rho\approx\rho_c$. One of our main results is that for
values  $\rho<\rho_c$, the pmf appears to scale similar to 
the distribution of cluster sizes in usual percolation 
\cite{stauffer1979,stauffer1994}, i.e.\
\begin{equation} 
n_\ell(\rho) \propto \ell^{-\tau} \exp\{-T_{\rm L}(\rho) \ell\} \quad \text{for} \quad \rho<\rho_c.\label{eq:loopDistrib}
\end{equation}
Therein, the exponential factor accounts for the observation that below
the critical point $\rho_c$ the proliferation of ``long'' (still non-spanning) loops is 
suppressed due to some finite ``loop size cut-off parameter''. The latter might be captured by means of a 
scaling parameter $T_{\rm L}(\rho)$ \cite{hindmarsch1995} which depends on the 
subtleties of the disorder. 
Its inverse $\ell_0(\rho)=1/T_{\rm L}(\rho)$ relates to
a typical length scale to which the perimeter of the loops is
limited at a given value of $\rho$ and it should not depend on the side length
$L$ of the system (at least in the limit of large system sizes where 
a loop of, say, length $\ell_0$ fits well into the simulation box).
Therefore, loop configurations a\cite{stauffer1979,stauffer1994} small values of $\rho$ are consistent
with a spanning probability $P_L(\rho\!<\!\rho_c)\rightarrow 0$ 
in the limit of large system sizes ($L\rightarrow\infty$).
As the critical point is approached from below 
the loop size cut-off parameter vanishes,
giving rise to the purely algebraic decay 
of $n_\ell$ observed at $\rho_c$, as in Eq.\ (\ref{eq:loopDistribCrit}), 
featuring loops with length $\ell$ on virtually all length scales. 
(A qualitatively similar observation in the context of high-$T_c$ superconductors is referred to 
as ``Onsager vortex-loop unbinding'' that signals the superconductor to normal metal
transition \cite{nguyen1998,nguyen1999}.
Further, in string theory, the analog observation is referred to as ``Hagedorn transition'' \cite{hindmarsch1995,antunes1998}.)
The decrease of the parameter $T_{\rm L}(\rho)$ can be related to a second, independent 
exponent that, in addition to $\tau$, serves to characterize the ensemble 
of small loops. The respective critical exponent $\sigma$ is defined via 
\begin{equation} 
T_{\rm L}(\rho) \propto |\rho-\rho_c|^{1/\sigma}~, \label{eq:lineTension}
\end{equation}
where $\sigma$ might be referred to as ``loop-size cut-off'' exponent
(i.e.\ the critical exponent related to the loop-size cut-off parameter $T_{\rm L}$)
and where $\rho$ approaches $\rho_c$ from below.
Similarly, the corresponding 
lengthscale $\ell_0$, to which the loops are confined, diverges. This 
implies that loops might get arbitrarily long, limited only by 
the finite size of the underlying lattice. One might expect a maximal 
loop length of 
$\ell_{\rm max}\!\sim\!L^{d_{f}}$, where $d_{f}$ denotes the fractal scaling dimension 
of the loops.
Thus, at $\rho_c$ and in the limit $L\to \infty$, the distribution of the loop perimeter 
exhibits an algebraic decay, solely governed by the fisher 
exponent $\tau$. 
Finally, according to scaling theory \cite{stauffer1979,stauffer1994}, the 
scaling relations
\begin{subequations}
\begin{eqnarray}
\nu_{\rm p} d_{f,{\rm p}} & =& 1/\sigma,  \label{eq:scaling:sigma}\\
\gamma_{\rm p} & = & (3-\tau)/\sigma  \label{eq:scaling:gamma}
\end{eqnarray}
\end{subequations}
 should hold, relating $\sigma$ 
(as measured 
for the small loops) to $\nu_{\rm p}$, $d_{f,{\rm p}}$ and $\gamma_{\rm p}$ 
(all measured from the system 
spanning loops; indicated by the subscript {\rm p}).
These three exponents signify 
the critical exponents that describe
the divergence of  the correlation length, the scaling dimension of the loops, 
and the fluctuations of the loop order-parameter,
 respectively.

The remainder of the present article is organized as follows.
In section \ref{sect:model}, we introduce the model in 
more detail and we outline the algorithm used compute the loop 
configurations. In section \ref{sect:results}, we list the results of 
our numerical simulations and in section \ref{sect:conclusions} we 
conclude with a summary.
Note that an extensive summary of this paper
is available at the \emph{papercore database} \cite{papercore}.

\section{Model and Algorithm\label{sect:model}}

In the present article we consider hypercubic lattice graphs 
$G\!=\!(V,E)$ with side length $L$ and 
fully periodic boundary conditions (BCs) for all relevant dimensions
$d=2$ through $6$.  
The considered graphs have $N\!=\!|V| \!=\!L^d$ sites
$i\!\in\!V$ and a number of $|E| \!=\!z N/2$ undirected edges  
$\{i,j\}\!\in\!E$ that join adjacent sites $i,j\!\in\!V$.  
Above, $z$ signifies the coordination number of the lattice geometry, where
$z=2d$.
We further assign a weight $\omega_{ij}$ to each $\{i,j\}\in E$. 
These weights represent quenched random variables that introduce disorder 
to the lattice.
Here we consider independent identically distributed 
weights which either have just weight one (probability 1-$\rho$)
or are drawn (probability $\rho$) 
from a Gaussian distribution with zeor mean and veriance one.
Hence, the disorder distribution is given by
\begin{equation}
P(\omega)=\rho \exp{(-\omega^2/2)}/\sqrt{2\pi} + (1-\rho) \delta(\omega-1), \label{eq:disorderDistrib}
\end{equation} 
that explicitly allows for loops $\mathcal{L}$ with a negative total weight
$\omega_{\mathcal{L}}\!=\!\sum_{\{i,j\}\in\mathcal{L}}\omega_{ij}$.
To support intuition: For any nonzero value of the disorder parameter $\rho$, a sufficiently
large lattice will exhibit at least ``small'' loops that exhibit a negative weight,
see Fig.\ \ref{fig:samples2D}(c). If the disorder parameter is large enough,
system spanning loops with negative weight will exist, 
see Figs.\ \ref{fig:samples2D}(a). 

The NWP problem then reads as follows:
Given a realization of the disorder for a hypercubic lattice graph $G$, determine a set 
$\mathcal{C}$ of loops such that the configuration energy, defined as the 
sum of all the loop-weights 
$\mathcal{E}\!=\!\sum_{\mathcal{L} \in \mathcal{C}} \omega_{\mathcal{L}}$, 
is minimized. As further optimization constraint, the loops are not
allowed to intersect.
Note that due to the ``energy minimization principle'' of the optimization 
procedure, the weight of an individual loop is necessarily smaller than zero. 
The configuration energy $\mathcal{E}$ is the quantity subject to
optimization and the result of the optimization procedure is a set of
loops $\mathcal{C}$, obtained using an appropriate transformation of
the original graph \cite{ahuja1993}.  
For the transformed graphs, \emph{minimum-weight perfect matchings} (MWPMs)
\cite{cook1999,opt-phys2001,melchertThesis2009} are calculated, that 
serve to identify the loops for a given realization of the disorder. 
Since exact MWPMs can be obtained in polynomial time,
 this procedure allows for an efficient implementation 
\cite{practicalGuide2009} of the  simulation algorithms.
Here, we give a brief description of the algorithmic procedure that yields a 
minimum-weight set of loops for a given realization of the disorder. 
Fig.\ \ref{fig2abcd} illustrates the three basic steps, detailed below:
\begin{figure}[t!]
\centerline{
\includegraphics[width=1.0\linewidth]{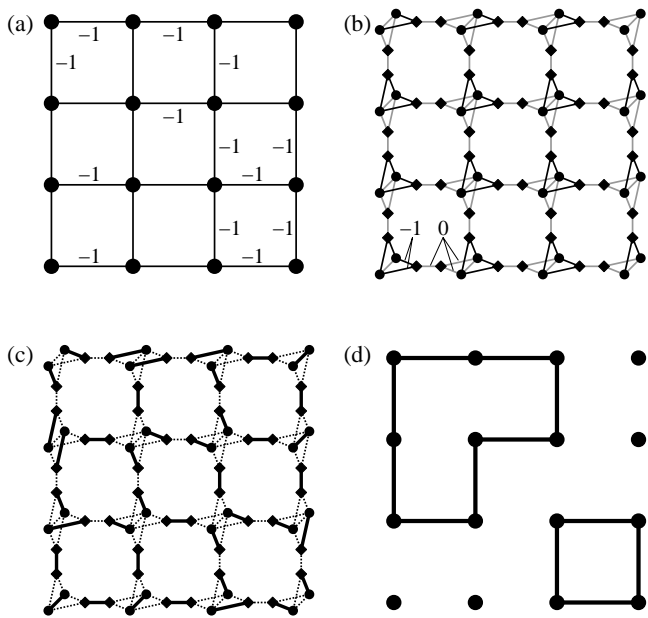}}
\caption{Illustration of the algorithmic procedure:
(a) original lattice $G$ with edge weights. 
For clarity, a bimodal distribution that yields edge-weights $\pm1$ is considered.
Further, only negative edge-weights are shown. Unlabeled edges have weight $+1$. 
(b) auxiliary graph $G_{\rm A}$ with proper weight assignment. Black 
edges carry the same weight as the respective edge in the original
graph and gray edges carry zero weight,
(c) minimum-weight perfect matching (MWPM) $M$: bold edges are matched 
and dashed edges are unmatched, and
(d) loop configuration (bold edges) that corresponds to the MWPM 
depicted in (c).
\label{fig2abcd}}
\end{figure}  

(1) each edge, joining adjacent sites on the original graph $G$,  is
replaced by a path of 3 edges.  Therefore, 2  ``additional'' sites
have to be introduced for each edge in $E$.  Therein, one of
the two edges connecting an additional site to an original site gets
the same weight as the corresponding edge in $G$. The remaining  two
edges get zero weight.  The original sites $i\in V$ are then
``duplicated'',  i.e. $i \rightarrow i_{1}, i_{2}$, along with all
their incident edges and the corresponding weights. 
 For each of these pairs of duplicated sites,
one additional  edge $\{i_1,i_2\}$ with zero weight is added that
connects the two sites $i_1$ and $i_2$.  The resulting auxiliary graph
$G_{{\rm A}}=(V_{{\rm A}},E_{{\rm A}})$  is shown
in Fig.\ \ref{fig2abcd}(b), where additional sites appear as squares and
duplicated  sites as circles. Fig.\ \ref{fig2abcd}(b) also illustrates
the weight  assignment on the transformed graph $G_{{\rm A}}$.  Note
that while the original graph (Fig.\ \ref{fig2abcd}(a)) is symmetric, the
transformed graph (Fig.\ \ref{fig2abcd}(b)) is not. This is due to the
details of the mapping procedure and the particular weight assignment
we have chosen.  A more extensive description of the mapping can be
found in \cite{melchert2007}.

(2) a MWPM on the auxiliary graph is determined via exact
combinatorial optimization algorithms \cite{comment_cookrohe}.  A MWPM
is a minimum-weighted subset $M$ of $E_{\rm A}$, such that
each site  contained in $V_{\rm A}$ is met by precisely one
edge in $M$.  This is illustrated in Fig.\ \ref{fig2abcd}(c), where the
solid edges  represent $M$ for the given weight assignment. The dashed
edges are  not matched.  Due to construction, the auxiliary graph
consists of an even number of sites and 
the transformation procedure described in step (1) guarantees
that a perfect matching exists.  
Note that a MWPM can be computed in polynomial time as a function
of the number of sites, hence large systems with hundreds of thousands
of sites are feasible.

(3) finally it is possible to find a relation between the matched
edges $M$  on $G_{\rm A}$ and a configuration of negative-weighted
loops  $\mathcal{C}$ on $G$ by  tracing back the steps of the
transformation (1). As regards this, note that each edge  contained
in $M$ that connects an additional site (square) to a duplicated  site
(circle) corresponds to an edge on $G$ that is part of a loop, see
Fig.\ \ref{fig2abcd}(d).
Note that, by construction of the auxiliary graph,
 for each site $i_1$ or $i_2$ matched in this
way, the corresponding ``twin'' site $i_2$/$i_1$ must be matched
to an additional site as well. This guarantees that wherever a path ``enters''
a site of the original graph, the paths also ``leaves'' the site, corresponding
to the defining condition of loops.
All the edges in $M$ that connect like sites (i.e.\ duplicated-duplicated, or
additional-additional)  carry zero weight and do not contribute to a
loop on $G$.  
Once all loop segments are found, a depth-first search \cite{ahuja1993,opt-phys2001} can  be used to
identify the loop set $\mathcal{C}$ and to determine the geometric 
properties of the individual loops. 
Here, the exemplary weight assignment illustrated 
in Fig.\ \ref{fig2abcd}(a) yields 2 loops, 
i.e.\ $\mathcal{C}\!=\!\{\mathcal{L}_1, \mathcal{L}_2\}$, with
weights $\omega_{\mathcal{L}_1}\!=\!\omega_{\mathcal{L}_2}\!=\!-4$
and lengths $\ell_1\!=\!\sum_{\{i,j\}\in\mathcal{L}_1}1\!=\!8$, 
$\ell_2\!=\!4$. Hence, the configurational energy reads $\mathcal{E}\!=\!-8$. 

The result of the calculation is a collection $\mathcal{C}$
of loops such that the total loop weight, and consequently the
configuration energy $\mathcal{E}$, is minimized. 
Hence, one obtains a global collective optimum of the system.  
Obviously, all loops that contribute to $\mathcal{C}$ possess a negative weight.  
Also note that the choice of the weight assignment in step (1) is not unique, 
i.e.\ there are different possibilities to choose a weight assignment
that all result in equivalent sets of matched edges on the transformed
lattice, corresponding to the minimum-weight collection of loops on
the original lattice. Some of these weight assignments result in a more
symmetric  transformed graph, see e.g. \cite{ahuja1993}. However, this
is only a technical issue that does not affect the resulting loop
configuration. Finally, for the purpose of illustration, a small $2D$ lattice graph with
free BCs was chosen intentionally. The algorithmic procedure extends 
to higher dimensions and fully periodic BCs in a straight-forward manner.

In the remainder of the article, we will use the procedure outlined above in
order to study the pmf of loop lengths in the NWP model on 
hypercubic lattice graphs in $d=2$ through $6$.

\section{Results \label{sect:results}}

Within our extensive numerical studies, we 
performed exact NWP loop calculations for dimensions $d=2$ through $6$
for various values $\rho\le \rho_c$ while averaging over
many realizations of the disorder. Details are given 
in Tab.\ \ref{tab:sim_parameters}.

%
\begin{table}[b!]
\caption{\label{tab:sim_parameters} 
Simulation parameters: we performed our study for
$n_\rho$ values of the disorder
parameter $\rho$ in intervals $[\rho_1,\rho_2]$  
and for a number $n_R$ of realizations,
for the different dimensions $d$ and system sizes $L$.}
\begin{ruledtabular}
\begin{tabular}[c]{llrrr}
$d$  & $L$ & $[\rho_1,\rho_2]$ & $n_\rho$ & $n_R$\\
\hline
2 & 256   & [0.24,\, 0.34]  & 11 &  $\sim 2\!\times\!10^4$\\
2 & 512   & [0.24,\, 0.34]  & 16 & $6400$ \\
3 & 64    & [0.075,\, 0.1245] & 100 & $4800$ \\
4 & 21    & [0.022,\, 0.058] & 19 & $8000$ \\
5 & 12    & [0.02,\, 0.038] & 19 & $6400$  \\
6 & 6     & [0.015,\, 0.025] & 21 & $\sim 5\!\times\!10^4$ 
\end{tabular}
\end{ruledtabular}
\end{table}

In order to get a grip on the loop-size cut-off parameter $\sigma$
for a particular hypercubic lattice setup of dimension $d$, 
the pmf $n_\ell(\rho)$ of the loop length needs to be obtained for different 
values of the disorder parameter $\rho\!\leq\!\rho_c$. 
Then, a best fit to the form $n_\ell(\rho)=n_0 \ell^{-\tau} \exp\{-T_{\rm L}(\rho) \ell\}$
might be used to obtain the three fit parameters $n_0$, $\tau$, and $T_{\rm L}(\rho)$ for different
values of $\rho$.
Finally, the sequence of fit parameters $T_{\rm L}(\rho)$ might be analyzed to yield the 
exponent $\sigma$ according to Eq.\ (\ref{eq:lineTension}).

However, a different procedure appears to be more appealing: from previous 
simulations in dimensions $d=2$ through $7$, reported in Refs.\ \cite{melchert2008,melchert2010a},
we found that the pmf $n_\ell(\rho_c)$ exhibits an algebraic decay 
governed by the exponent $\tau$. 
For the largest lattice graphs simulated for the various dimensions $d$, we obtain
the numerical estimates listed in Tab.\ \ref{tab:tab1}.
For the corresponding data analyses, very small loops have to be neglected since
they are affected by the granularity of the lattice and very large loops have 
to be withdrawn since they are affected by the lattice boundaries.
Once the exponent $\tau$ for a given dimension $d$ is obtained, it can be
utilized in the analysis of the loop length pmf at $\rho<\rho_c$
to limit the number of fit parameters to only two (i.e.\ $n_0$ and $T_{\rm L}$), allowing for a more precise
estimate of the individual values of $T_{\rm L}(\rho)$.

\begin{figure}[t!]
\centerline{
\includegraphics[width=1.0\linewidth]{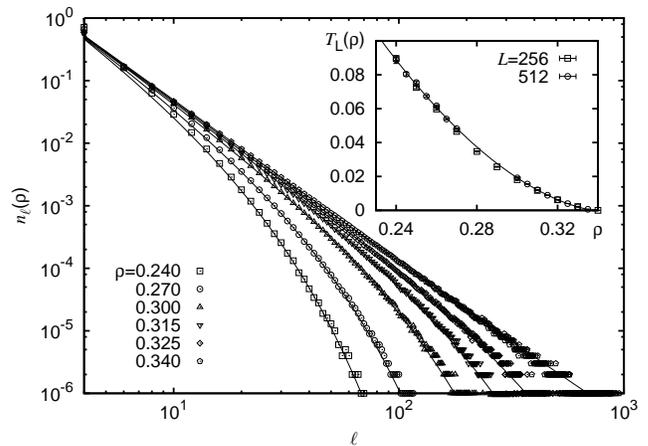}}
\caption{Results for $d=2$ square lattice graphs.
The main plot shows the probability mass functions $n_\ell(\rho)$ of the loop perimeter 
$\ell$ for different values of the disorder parameter $\rho$ for 
square systems of side length $L=512$. The data curves illustrate the
suppression of loops with a perimeter larger than some 
cut-off length scale $\ell_0$, related to a finite ``loop size cut-off parameter'' $T_{\rm L}(\rho)$ 
for $\rho<\rho_c$ (see text). The data curves fit well to functions
of the form of Eq.\ (\ref{eq:loopDistrib}). Right at $\rho_c$ the data curve
exhibits a pure algebraic decay according to Eq.\ (\ref{eq:loopDistribCrit}).
The inset compares the resulting values of the cut-off parameter $T_{\rm L}(\rho)$
for systems of side length $L=512$ and $L=256$, where the solid line indicates a best 
fit to the $L=512$ data using a function with four free parameters as explained in the text.\label{fig:pmf2D}} 
\end{figure}

So as to get a grip on the loop size cut-off parameter $\sigma$, 
the loop perimeter distribution was obtained for different 
values of the disorder parameter $\rho\!\leq\!\rho_c$ and
the data was fitted using a function as given in
 Eq.\ (\ref{eq:loopDistrib}), with an additional normalization factor.
E.g., in $d=2$ we use $\tau\!=\!2.59$ fixed, 
see Tab.\ \ref{tab:tab1} and Fig.\ \ref{fig:pmf2D},
wherein the fit intervals for individual values of $\rho$ 
where restricted to the range $[\ell_{\rm min},\ell_{\rm max}]$.
We further fixed $\ell_{\rm min}=10$ and $\ell_{\rm max}=60$ at $\rho=0.24$
(the upper bound shifting up to $\ell_{\rm max}=150$ at $\rho_c$).
The resulting values for the cut-off parameter were then analyzed using 
a three parameter fit to functions of 
the form of Eq.\ (\ref{eq:lineTension}), i.e.\ 
$T_{\rm L}(\rho) = A |\rho-\rho_c^{\rm eff}|^{1/\sigma}$.
Therein the amplitudes $A$ are not of interest and the 
effective critical points $\rho_c^{\rm eff}$ can be expected to 
differ slightly from the asymptotic critical points listed in 
Tab.\ \ref{tab:tab1} (see discussion below). The resulting loop-length cut-off parameters $\sigma$
are also listed in Tab.\ \ref{tab:tab1}.

\begin{figure}[t!]
\centerline{
\includegraphics[width=1.0\linewidth]{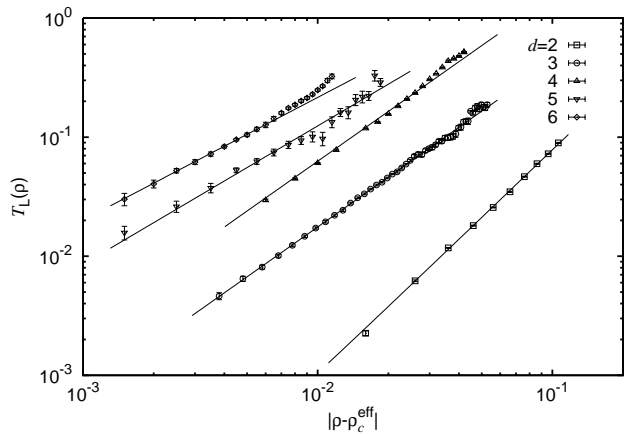}}
\caption{Scaling analysis of the loop size cut-off parameter $T_{\rm L}(\rho)$ in dimensions
$d=2$ through $6$.
The solid lines indicate fits to functions of the form 
$T_{\rm L}(\rho) = A |\rho-\rho_c^{\rm eff}|^{1/\sigma}$.
Therein the amplitudes $A$ are not of interest and the 
effective critical points $\rho_c^{\rm eff}$ can be expected to 
differ slightly, but within statistical error bars,
 from the asymptotic critical points listed in 
Tab.\ \ref{tab:tab1}. The resulting loop-length cut-off parameters $\sigma$
are also listed in Tab.\ \ref{tab:tab1}. \label{fig:lineTension}}
\end{figure}

For $d=2$ systems of side length $L=512$ the analysis yields
$\sigma=0.53(3)$ and $\rho_c^{\rm eff}=0.344(2)$,
see Fig.\ \ref{fig:lineTension}
(at $L=256$ we find $\rho_c^{\rm eff}\!=\!0.346(2)$ and $\sigma=0.53(2)$). 
For all values of $\rho$ considered, the data curves of $T_{\rm L}$ at $L=256$ and
$L=512$ compare well as shown in the inset of Fig.\ \ref{fig:pmf2D}.
This finding is further consistent with the usual scaling relation 
$1/(\nu_{\rm p} d_{f,{\rm p}})\!=\!\sigma$ that relates $\sigma$ to $\nu_{\rm p}$ and $d_{f,{\rm p}}$,
where the latter two exponents signify 
the critical exponent that describes the divergence of 
the correlation length and the scaling dimension of the loops at 
the critical point, respectively. From the respective values 
previously obtained \cite{melchert2008}
one readily finds $1/(\nu_{\rm p} d_{f,{\rm p}})\!=\!0.53(3)$.
Only the location of the (effective) critical point $\rho_c^{\rm eff}\!=\!0.344(2)$, as 
estimated from the complete ensemble of loops at the particular system size $L=512$, differs slightly 
from the one obtained from the scaling analysis of the 
percolating loops, i.e.\ $\rho_c\!=\!0.340(1)$ \cite{melchert2008} 
(a similar effect was found in the analysis of a $d=3$ vortex loop network in Ref.\ \cite{kajantie2000}). 
Bearing in mind that the latter value represents an extrapolation
to the thermodynamic limit, the aforementioned difference is likely 
due to finite-size effects. 
In this regard it does not come as a surprise that an analysis 
of the loop size cut-off parameter according to Eq.\ (\ref{eq:lineTension})
for fixed $\rho_c^{\rm eff}=0.340$ 
(instead of $\rho_c^{\rm eff}=0.344$) yields the 
exponent $\sigma=0.61(1)$ which significantly overestimates the result obtained 
from the three parameter fit reported above. 
Finally, a four parameter fit according to 
$T_{\rm L}(\rho) = T_{\rm  L}^\prime + A |\rho-\rho_c^{\rm eff}|^{1/\sigma}$ at $L=512$
(see inset of Fig.\ \ref{fig:pmf2D})
results in the estimates $T_{\rm L}^\prime=0.001(1)$, $\rho_c^{\rm eff}=0.348(6)$ and $\sigma=0.55(3)$, with 
the latter two fit parameters in agreement with the ones found above and $T_{\rm L}^\prime$
in agreement with zero as one would naively expect.

So as to facilitate a qualitative comparison, considering $d=3$ hypercubic lattices with $L=64$ we obtained the effective critical 
point $\rho_c^{\rm eff}=0.1278(1)$ which slightly overestimates the asymptotic value of
$\rho_c=0.1273(3)$ (similar to what we observed above in $2D$), see Fig.\ \ref{fig:lineTension}. The scaling exponent $\sigma=0.71(1)$ compares well to the 
product $\nu_{\rm p} \cdot d_{f,{\rm p}} = 0.69(2)$ for the 
respective dimension. 
In the analysis of the data for $d>3$ we obtained most satisfactory fits by fixing the parameters
$\rho_c^{\rm eff}$ to their expected asymptotic values $\rho_c$ listed in Tab.\ \ref{tab:tab1}.
The analysis of the loop size cut-off parameter in dimensions $d=2$ through
$6$ are shown in Fig.\ \ref{fig:lineTension} and summarized 
in Tab.\ \ref{tab:tab1}.

In higher dimensions we also checked that the loop size cut-off parameter $T_{\rm L}(\rho)$ at a given value
of $\rho$ is practically independent of the system size (however, for larger system sizes
the loop yield is bigger and hence the statistics get more reliable). 
E.g., for $d=4$ hypercubic lattices
at $\rho=0.026$, i.e.\ with some distance to the critical point $\rho_c=0.0640(2)$,
we obtained $T_{\rm L}(0.026)=0.43(5)$ at $L=16$ and $T_{\rm L}(0.026)=0.45(4)$ at $L=21$.
Close to the critical point at $\rho=0.058$ we further find $T_{\rm L}(0.058)=0.35(1)$ at $L=16$ and 
$T_{\rm L}(0.058)=0.036(1)$ at $L=21$.
Further, we observe that in any dimension considered, the scaling relation $\gamma_{\rm p}=(3-\tau)/\sigma$ appears 
to be satisfied within errorbars.

\section{Conclusions \label{sect:conclusions}}
In the presented analysis of the NWP model, we performed
numerical simulations on hypercubic lattice graphs for all
dimensions relevant for the model, i.e.\
$d\!=\!2$ through $6$.  The aim of the study was to characterize the
ensemble of small loops in the NWP model using two independent
critical exponents: the Fisher exponent $\tau$ (which was already
known from previous studies, see Ref.\ \cite{melchert2010a}), and the
loop-length cut-off exponent $\sigma$.  Both exponents can be
determined by means of an analysis of the probability mass function
$n_\ell(\rho)$ measuring the distribution of loop lengths $\ell$, 
considering a
sequence of different values $\rho$ close to but below the critical
point. This implies a huge numerical effort, since we had
to study in different dimensions large systems, for several
values of the disorder parameter, while averaging over many realizations
of the disorder.
 For the numerical simulations we used a mapping of the NWP
model to a combinatorial optimization problem that allows to obtain
configurations of minimum weight loops via exact algorithms.  Note
that due to the small side lengths of the lattice graphs that are
accessible in high dimensions,  the data analysis is notoriously
difficult at large values of $d$.  However, we find the results
regarding the exponent $\sigma$ consistent with the scaling relations
Eqs.\ (\ref{eq:scaling:sigma}) and (\ref{eq:scaling:gamma})
for any dimension considered (see Tab.\ \ref{tab:tab1}).
Thus, via this extensive numerical study we have completed a 
comprehensive description of the static behavior of the NWP model in all relevant dimensions
$d=2,\ldots,6$.

In particular, many of the $d=3$ loop models studied in the
 literature report on values for the  critical exponents $\tau$ and
 $\sigma$ that are close to $\tau=3.07(1)$ and $\sigma=0.71(1)$
 ($1/\sigma=1.41(2)$) found here. E.g.,
Ref.\ \cite{nguyen1999} obtains $\tau=2.4(1)$ (however, in a previous
study they report $\tau=3$, see Ref.\ \cite{nguyen1998}) and
$1/\sigma=1.45(5)$ (in that study the latter quantity was called $\gamma$) 
for the $d=3$ uniformly frustrated XY model
as well as for the  lattice Ginzburg-Landau model in a frozen gauge
approximation, and Ref.\ \cite{pfeiffer2002}  yields $\tau=2.8(1)$ and
$\sigma=0.6(1)$ for the strongly screened vortex glass model.  Note
that in $d=3$ all the critical exponents of the NWP problem appear to
be quite close to those  that describe the strongly screened vortex
glass model analyzed in Ref.\ \cite{pfeiffer2002}.  As regards this,
it appears to be tempting to conclude that in $d=3$ both models are in
the same universality class.


\begin{acknowledgments}
LA  acknowledges a scholarship of the German academic exchange service DAAD
within the ``Research Internships in Science and Engineering'' (RISE) program
and the City College Fellowship program for further support.
OM acknowledges financial support from the DFG (\emph{Deutsche Forschungsgemeinschaft}) 
under grant HA3169/3-1. 
The simulations were performed at the HPC Cluster HERO, located at 
the University of Oldenburg (Germany) and funded by the DFG through
its Major Instrumentation Programme (INST 184/108-1 FUGG) and the
Ministry of Science and Culture (MWK) of the Lower Saxony State and
at the GOLEM I cluster for scientific computing, also located at 
the University of Oldenburg.
\end{acknowledgments}


\bibliography{nwp_loopDistrib.bib}

\end{document}